\documentclass[11pt]{article}

\usepackage{graphicx}        % standard LaTeX graphics tool

\newcommand{\be}{\begin{equation}}
\newcommand{\ee}{\end{equation}}
\newcommand{\bea}{\begin{eqnarray}}
\newcommand{\eea}{\end{eqnarray}}
\newcommand{\ba}{\begin{eqnarray*}}
\newcommand{\ea}{\end{eqnarray*}}

\newcommand{\rd}{{\rm d}}

\newcommand{\ep}{\varepsilon}

\newcommand{\Fh}[2]{\,{}_#1F_#2}
\newcommand{\Fs}[3]{\!\!\left[\begin{array}{c}#1\,;\\#2\,;\end{array}#3\right]}

\newcommand{\Fot}[2]{\Fs{#1}{#2}{- \frac13}}

\newcommand{\Fmx}[2]{\Fs{#1}{#2}{1-x}}

\newcommand{\FX}[2]{\Fs{#1}{#2}{\frac{(1-9z)^2}{(1+3z)^3}}}
\newcommand{\FR}[2]{\Fs{#1}{#2}{\frac{27(1-z)^2z}{(1+3z)^3}}}

\newcommand{\Fus}[2]{\Fs{#1}{#2}{\frac{4tz+1-z-L}{4tz+1-z+L }}}

\begin{document}

\begin{titlepage}

\begin{flushright}
%\today  \\
DESY-06-032\\
SFB/CPP-06-13\\
\end{flushright}

\vspace*{0.2cm}
\begin{center}
{\Large {\bf Hypergeometric representation of the two-loop\\
equal mass sunrise diagram 
}}\\[2 cm]
{\bf  O.V.~Tarasov}
\footnote{On leave of absence from JINR,
141980 Dubna (Moscow Region), Russian Federation.}\\[1cm]

 {\it Deutsches Elektronen - Synchrotron DESY\\
       Platanenallee 6, D-15738 Zeuthen, Germany\\
     E-mail: {\tt Oleg.Tarasov@desy.de}}\\

\end{center}

\vspace*{1.0cm}

\begin{abstract}
A recurrence relation between equal mass two-loop sunrise
diagrams differing in dimensionality by 2 is derived and it's 
solution in terms  of Gauss' $\left._2F_1\right.$ and 
Appell's $F_2$ hypergeometric  functions is presented. For 
arbitrary space-time dimension $d$ the imaginary part of the diagram on 
the cut is found to be the $\left._2F_1\right.$  hypergeometric function 
with  argument proportional to the maximum  of the Kibble cubic form. 
The analytic expression for the threshold value of the diagram   in terms of  
the hypergeometric function  $\left._3F_2\right.$ of argument $-1/3$ 
is given.

\end{abstract}

\end{titlepage}

\newpage
\section{Introduction}

The evaluation of radiative corrections for  modern
high precision particle physics  is becoming a more and 
more demanding task.  Without inventing new mathematical methods
and new computer algorithms the progress in  calculating multi-loop, 
multi-leg Feynman diagrams depending on several momentum 
and mass scales will be not possible.

An important class of radiative corrections comes from 
self-energy type of  Feynman diagrams, which  also occur 
in evaluating  vertex, box and higher multileg  diagrams.
At the two-loop level different approaches for calculating 
self-energy diagrams are available \cite{2loopSelfEnergies}.
A general algorithm for reduction of  propagator type of
diagrams  to a minimal set of master integrals
was proposed in \cite{Tarasov:1997kx}.
This recurrence relations algorithm has been implemented in 
computer packages in \cite{Mertig:1998vk} and 
\cite{Martin:2003qz},\cite{Martin:2005qm}.
At  present   the most  advanced package available
for  calculating two-loop self-energy  diagrams with arbitrary 
massive particles was written by Stephen Martin and David
Robertson \cite{Martin:2005qm}.
It includes  procedures for numerical evaluation of master 
integrals with arbitrary masses and also
a database of analytically known master integrals.
Integral representation for master integrals with arbitrary masses 
in four dimensional space time was proposed in  \cite{Bauberger:1994hx}.

Despite intensive efforts by many authors not all two-loop 
self-energy integrals with a mass are known analytically.
Even the imaginary part of the simplest sunrise self-energy
diagram with three equal mass propagators  was not known 
for arbitrary space-time dimension $d$ until now. 
The two-loop sunrise integral with three equal masses was
investigated in many publications 
\cite{Mendels}- 
%\cite{Broadhurst:1987ei}
%\cite{Ford:pn},
%\cite{Davydychev:1992mt},
%\cite{Broadhurst:1993mw},
%\cite{Berends:1993ee}
%\cite{Fleischer:1998dw},
%\cite{Davydychev:1999ic},
\cite{Kniehl:2005bc}.
Small and large momentum expansions of this integral for
arbitrary space-time dimension $d$ can be found in 
\cite{Broadhurst:1993mw}. It's threshold expansion was
given in \cite{Davydychev:1999ic}.
A numerical procedure for evaluating the sunrise integral
was described in \cite{Fleischer:1998dw}. The very latest effort  
of an analytic 
calculation of the diagram, by using the differential equation
approach \cite{Kotikov:1990kg}, was undertaken in Ref. \cite{Laporta:2004rb}.

It is the purpose of this paper to describe a new method 
and to present the analytic result  for the  equal mass 
two-loop sunrise master integral. 
To accomplish our goal we use  the method of evaluation of 
master integrals by dimensional recurrences proposed in
\cite{Tarasov:1996br}. The application of this 
method to one-loop integrals was presented in \cite{Tarasov:2000sf}
and  \cite{Fleischer:2003rm}.  In the present publication we 
extend the method to two-loop integrals.

As was already discovered in the one-loop case, the solutions
of dimensional recurrences are combinations of hypergeometric
functions.  The knowledge of the hypergeometric representation of
an integral means that we possess the most complete mathematical 
information  available. This information
can be effectively used in several respects. First, through analytic
continuation formulae, the hypergeometric functions valid in one
kinematic domain can be re-expressed in a different kinematic 
region.
Second, these hypergeometric functions often have 
integral representation themselves, in which an expansion in
$\ep=(4-d)/2$ can be made, yielding expressions in logarithms,
dilogarithms, elliptic integrals, etc.. Since very similar 
hypergeometric functions come from  different kind of Feynman 
integrals the $\ep$ expansion derived in solving one problem can
be used in other applications.
Essential progress in the $\ep$ expansion of hypergeometric 
functions encountered in  evaluating Feynman diagrams was
achieved in \cite{Broadhurst:1993mw},\cite{Fleischer:1998dw},
\cite{EpsExpansion} and \cite{Kalmykov:2006pu}.
Third, because the hypergeometric series is convergent and well 
behaved in a particular region of kinematical variables, 
it can be numerically evaluated \cite{BBBB}, \cite{lsjk}.
In addition a hypergeometric representation allows an asymptotic
expansion of the integral in terms of ratios of different Gram
determinants or ratios of momentum and mass scales which can
provide fast numerical convergence of the result.

Our paper is organized as follows. In Sec. 2, 
we present the relevant difference equation connecting sunrise integrals
with dimensionality differing by 2 as well as the
differential equation for this integral.
In Sec. 3. the method for finding the full solution of the 
dimensional recurrency is elaborated. 
Explicit expression for the sunrise integral in terms of Appell's
function $F_2$ and Gauss' hypergeometric function $\left._2F_1\right.$
is constructed in  Sec. 4  and  in Sec. 5  the differential equation
approach and the method of dimensional recurrences is compared.
In the Appendix some useful formulae for the hypergeometric
functions $\left._2F_1\right.$, $F_1$ and $F_2$ are given together
with their integral representations.

\section{Difference and differential equations for the
  sunrise integral}

The generic two-loop self-energy type diagram in $d$ dimensional Minkowski
space with three equal mass  propagators  is given by the following integral:
\begin{equation}
J_3^{(d)}(\nu_1,\nu_2,\nu_3)
\equiv\int\!\!\int\frac{\rd^dk_1\rd^dk_2}{(i \pi^{d/2})^2}
\frac{1}{(k_1^2-m^2)^{\nu_1}
        ((k_1-k_2)^2-m^2)^{\nu_2}
	((k_2-q)^2-m^2)^{\nu_3} }.
\label{J3def}
\end{equation}
For integer values of $\nu_j$ the integrals (\ref{J3def})
can be expressed in terms of only three  basis integrals
$J_3^{(d)}(1,1,1)$, $J_3^{(d)}(2,1,1)$
and $J_3^{(d)}(0,1,1)= (T_1^{(d)}(m^2))^2 $  where
%%%%%%%%%%%%%%%%%%%%%%
\begin{equation}
T_1^{(d)}(m^2) 
= \int \frac{d^dk}{[i \pi^{\frac{d}{2}}]} \frac{1}{k^2-m^2}
~=~-\Gamma\left(1-\frac{d}{2}\right) m^{d-2}.
\end{equation}
The relation connecting  $d-2$ and $d$ dimensional integrals
$J_3^{(d)}(\nu_1,\nu_2,\nu_3)$ follows from the
relationship given in Ref. \cite{Tarasov:1997kx}:
\begin{eqnarray}
J_3^{(d-2)}(\nu_1,\nu_2,\nu_3)
&=& \nu_1 \nu_2  J_3^{(d)}(\nu_1+1,\nu_2+1,\nu_3)
\nonumber \\
&+& \nu_1 \nu_3 
J_3^{(d)}(\nu_1+1,\nu_2,\nu_3+1)
+ \nu_2 \nu_3 
J_3^{(d)}(\nu_1,\nu_2+1,\nu_3+1).
\label{dm2}
\end{eqnarray}
Relation (\ref{dm2}) taken at $\nu_1=\nu_2=\nu_3=1$ and
$\nu_1=2,~\nu_2=\nu_3=1$ gives two equations. To simplify these equations 
we use  the recurrence relations proposed in \cite{Tarasov:1997kx}.
From these two equations by shifting $d \rightarrow d+2$ 
two more relations follow. They are used to exclude
$J_3^{(d)}(2,1,1)$ from one of the relations, so that we  obtain
a difference equation for the master integral 
$J_3^{(d)}(1,1,1) \equiv J_3^{(d)}$:
%%%%%%%%%%%%%%%%%%%%%%
\begin{eqnarray}
12 z^3 (d+1) (d-1) (3 d+4) (3 d+2) &&J_3^{(d+4)}
\nonumber 
\\
-4 m^4(1-3 z) (1-42 z+9 z^2) z (d-1) d &&J_3^{(d+2)}
\nonumber 
\\
-4 m^8(1-z)^2 (1-9 z)^2 &&J_3^{(d)}
\nonumber \\
= 3 z [(z+1) (27 z^2+18 z-1) d^2-4 z (1+9 z) d-48 z^2]
m^{2d+2} &&\Gamma\left(-\frac{d}{2} \right)^2, 
\label{j3difference}
\end{eqnarray}
where
\begin{equation}
z=\frac{m^2}{q^2}.
\end{equation}

The integral $J_3^{(d)}$ satisfies also a second order
differential equation  \cite{Broadhurst:1993mw}. Taking the second 
derivative of $J_3^{(d)}$ with respect to mass gives
\begin{equation}
\frac{ d^2}{ dm^2~ d m^2}
J_3^{(d)}(1,1,1) = 6 J_3^{(d)}(2,2,1)+ 6 J_3^{(d)}(3,1,1).
\label{secondDerivative}
\end{equation}
Again using the recurrence  relations from \cite{Tarasov:1997kx}, the 
integrals on the r.h.s  can be reduced to the same three basis integrals.
%$J_3^{(d)}(1,1,1) $ , $J_3^{(d)}(2,1,1)$ and
% $J_3^{(d)}(0,1,1)=(T_1^{(d)}(m^2))^2$.
Using
\begin{equation}
J_3^{(d)}(2,1,1)= \frac13~\frac{d}{ d m^2}
  J_3^{(d)}(1,1,1)
\end{equation}
from (\ref{secondDerivative}) we obtain:
\begin{eqnarray}
  2(1-z)(1-9z)z^2 \frac{{ d^2}J_3^{(d)}}{{ d}z^2}
% \nonumber \\
 &-& z[9 z^2 (d-4)+10 z (d-2)  +8-3 d]
 \frac{{ d}J_3^{(d)}}{{ d} z}
 \nonumber \\
 &+& (d-3)[z(d+4)+d-4 ] J_3^{(d)}
% \nonumber \\
 = 12z m^{(2d-6)} \Gamma^2\left(2-\frac{d}{2}\right).
\label{difequation}
\end{eqnarray}
The differential equation (\ref{difequation}) will be used in  
Sec.4  to find the momentum dependence of arbitrary periodic constants 
in the solution of the difference equation (\ref{j3difference}).

\section{Solution of the dimensional recurrency}

Equation (\ref{j3difference}) is a second order inhomogeneous 
equation with polynomial coefficients in $d$.
The full solution of this equation  is given by (see
Ref. \cite{Milne-Thomson} and references therein):
\begin{equation}
J_3^{(d)}= J_{3p}^{(d)} + \widetilde{w}_a(d)J_{3a}^{(d)}
                        + \widetilde{w}_b(d)J_{3b}^{(d)},
\label{FullSolution}
\end{equation}
where $J_{3p}^{(d)}$ is a particular solution of
(\ref{j3difference}), $J_{3a}^{(d)},
J_{3b}^{(d)}$ is a fundamental system of solutions of the
associated homogeneous equation and $\widetilde{w}_a(d)$,
$\widetilde{w}_b(d)$ are arbitrary periodic functions of $d$
satisfying relations:
\begin{equation}
\widetilde{w}_a(d+2)=\widetilde{w}_a(d),
~~~~~~~~~~\widetilde{w}_b(d+2)=\widetilde{w}_b(d).
\end{equation}

The  order of the polynomials in $d$ of the associated 
homogeneous difference equation can  be reduced by making
the substitution
\begin{equation}
J_3^{(d)} = \frac{\Gamma\left(\frac{d-2}{2}\right)}
{\Gamma\left(\frac{3d}{2}-3\right)\Gamma\left(\frac{d-1}{2} \right)}
\overline{J}_3^{(d)}.
\label{J3bar}
\end{equation}
The associated homogeneous equation for $\overline{J}_3^{(d)}$ takes
the simpler form
\begin{eqnarray}
&&  \frac{16 z^3}{27 m^8 (1-z)^2 (1-9z)^2}
  ~\overline{J}_3^{(d+4)}
\nonumber \\
&&~~~~~~~~~~~~~
- \frac{2 d (1-3z)(1-42z+9z^2)z}{27 m^4(1-z)^2 (1-9z)^2}
  ~\overline{J}_3^{(d+2)}
- \frac{(3d-2)(3d-4)}{36}~ \overline{J}_3^{(d)}=0.
\label{j3bar}
\end{eqnarray}
Putting 
\begin{equation}
d=2k-2\varepsilon,~~~~ y^{(k)}=\rho^{-k}~\overline{J}_3^{(2k-2\varepsilon)},
\label{Yk}
\end{equation}
we transform Eq.(\ref{j3bar}) to a standard form
\begin{equation}
A \rho^2 y^{(k+2)}+(B+C~k) \rho y^{(k+1)} - (\alpha+k)(\beta+k)y^{(k)}=0,
\label{sform}
\end{equation}
where
\begin{eqnarray}
A&=&\frac{16z^3}{27m^8(1-z)^2(1-9z)^2},~~~~
B= \frac{4\varepsilon}{27}
\frac{(1-3z)(1-42z+9z^2)z}{m^4(1-z)^2(1-9z)^2},
\nonumber \\
C&=& -\frac{B}{\varepsilon},
~~~~~~\alpha=-\varepsilon-\frac13,~~~~~~\beta=-\varepsilon -\frac23,
\end{eqnarray}
and $\rho$ is for the time being, an arbitrary constant. 
In order to get Eq.(\ref{sform}) into a more convenient form,
we will define three parameters $\rho$, $x$ and $\gamma$ by
the equations
\begin{equation}
A\rho^2=x(1-x),~~B\rho=\gamma-(\alpha+\beta+1)x,~~C\rho=1-2x.
\end{equation}
These have the solution
\begin{eqnarray}
\label{X}
x&=&\frac{1-2C\rho}{2}=\frac{(1-9z)^2}{(1+3z)^3}~=~
\frac{q^2(q^2-9m^2)^2}{(q^2+3m^2)^3},
 \\
\nonumber \\
\rho&=&\frac{1}{\sqrt{4A+C^2}}= 
\frac{27}{4}\frac{m^4(1-z)^2(1-9z)^2}{z(1+3z)^3}~=~
\frac{27}{4} \frac{m^2(q^2-m^2)^2(q^2-9m^2)^2}{(q^2+3m^2)^3},
 \\
\nonumber \\
\gamma&=&B\rho+(\alpha+\beta+1)x = -\varepsilon,
\label{XandRO}
\end{eqnarray}
and Eq. (\ref{sform}) can accordingly be written in the form
\begin{equation}
x(1-x)y^{(k+2)}+[(1-2x)k + \gamma -(\alpha+\beta+1)x]y^{(k+1)}
-(\alpha+k)(\beta+k)y^{(k)}=0.
\label{Ykequ}
\end{equation}
The fundamental system of solutions of this  equation consists
of two hypergeometric functions \cite{Milne-Thomson}. For example,
in the case when $|1-x|<1$ (large $q^2$) the solutions are
\begin{eqnarray}
y_1^{(k)}&=&(-1)^k \frac{\Gamma(\alpha+k) \Gamma(\beta+k)}
           {\Gamma(\alpha+\beta-\gamma+k+1)}
	   ~\left._2F_1(\alpha+k,
	   \beta+k,\alpha+\beta-\gamma+k+1 ;1-x),\right.
\nonumber \\
y_2^{(k)}&=& \frac{\Gamma(\alpha+\beta-\gamma+k)}
{(1-x)^k}
\left._2F_1(\gamma-\alpha,
	    \gamma-\beta,
	    \gamma-\alpha-\beta+1-k ;1-x).\right.
\label{nearX1}
\end{eqnarray}
Once we know the solutions of the homogeneous equation,
a particular solution  $J_{3p}^{(d)}$ can be obtained by
using Lagrange's method of variation of parameters.
Lagrange method for finding a particular solution is well
described in \cite{Milne-Thomson}.
The application of the method is straightforward but
tedious. Explicit result will be given in the next section.

It is interesting to note that the argument of the 
Gauss' hypergeometric function is related to the maximum
of the Kibble cubic form~\cite{Kibble}:

\begin{equation}
\Phi(s,t,u)  = stu - (s+t+u) m^2(m^2  + q^2) + 2 m^4(m^2 + 3q^2),
\end{equation}
provided that the following condition  is satisfied:
\begin{equation}
\label{s+t+u}
s+t+u = q^2 + 3m^2.
\end{equation}
The maximal value
$\Phi_{\rm max}=\frac{1}{27}\;q^2(q^2-9m^2)^2$ occurs at
$s=t=u=\frac{1}{3}\;(q^2+3m^2)$ and we see that the
kinematical variable (\ref{X} ) can be written as
\begin{equation}
x=\frac{\Phi(s,t,u)}{stu}\left|_{s=t=u=\frac{1}{3}\;(q^2+3m^2)}
\right..
\end{equation}
This observation may be useful in finding
the characteristic variable in the general mass case
\cite{Davydychev:2003cw}. Also one can try to apply the
method described above to find the imaginary part of the
sunrise integral in the general mass case in arbitrary
space-time dimension.

%%%%%%%%%%%%%%%%%%%%%%%%%%%%%%%%%%%%%%%%%%%%%%%%%%%%%%%%%%

\section{Explicit analytic expression for $J_3^{(d)}$}

%%%%%%%%%%%%%%%%%%%%%%%%%%%%%%%%%%%%%%%%%%%%%%%%%%%%%%%%%%%%
To find the full solution of Eq. (\ref{j3difference}) we assume
that $q^2$ is large.
The region of large momentum squared corresponds to  $x\sim 1$ 
and therefore as a fundamental system of solutions of the homogeneous 
equation we take  $y_1^{(k)}$ and  $y_2^{(k)}$. According to 
(\ref{J3bar}), (\ref{Yk}) and (\ref{nearX1}) the solution of the 
associated  homogeneous difference equation  will be of the form
\begin{eqnarray}
J^{(d)}_{3,h}&=&w_1(z)\frac{\Gamma\left(\frac{d}{2}-\frac13\right)
\Gamma\left(\frac{d}{2}-\frac23\right)
\Gamma\left(\frac{d-2}{2}\right)}
{\Gamma\left(\frac{d}{2}\right)\Gamma\left(\frac{3d}{2}-3\right)
\Gamma\left(\frac{d-1}{2}\right)}
\rho^{\frac{d}{2}}e^{i\pi \frac{d}{2}}
 \Fh21\Fmx{\frac{d}{2}-\frac13,\frac{d}{2}-\frac23}{\frac{d}{2}}
\nonumber \\
&+&w_2(z)\frac{\Gamma^2\left(\frac{d-2}{2}\right)}
{\Gamma\left(\frac{3d}{2}-3\right) 
 \Gamma\left(\frac{d-1}{2}\right)}\frac{\rho^{\frac{d}{2}}}
 {(1-x)^{\frac{d}{2}}}
 \Fh21\Fmx{\frac13,\frac23}{2-\frac{d}{2}}.
\label{homoLarge}
\end{eqnarray}
The arbitrary periodic functions $w_1(z)$ and $w_2(z)$ 
can be determined either from the  $d\rightarrow \infty$ 
asymptotics or using the differential equation (\ref{difequation}). 
Substituting (\ref{homoLarge}) into (\ref{difequation})  we obtain 
two simple equations
\begin{eqnarray}
&&z(1-z)(1+3z)(1-9z)\frac{dw_1(z)}{dz}-2(1+6z-39z^2)w_1(z)=0,
\nonumber \\
&&z(1+3z)(1-9z)\frac{dw_2(z)}{dz}+3(1-z)w_2(z)=0.
\label{difur_w1w2}
\end{eqnarray}
Both equations are independent of $d$ and their solutions
\begin{equation}
w_1(z)=\frac{\kappa_1 z^2(1+3z)^2}{(1-9z)^2(1-z)^2},
~~~w_2(z)=\frac{\kappa_2 z^3}{(1+3z)(1-9z)^2},
\label{w1w2}
\end{equation}
determine the periodic functions up to integration constants  
$\kappa_1,~\kappa_2$  which we fix from the first two terms 
of the large momentum expansion of $J_3^{(d)}$ 
presented in \cite{Broadhurst:1993mw}:
\begin{equation}
J_3^{(d)} = m^{2-4\ep}\Gamma^2(1+\ep)
\left[\frac{\Gamma(-1+2\ep)\Gamma^3(1-\ep)}
{z\Gamma^2(1+\ep)\Gamma(3-3\ep)}(-z)^{2\ep}
+\frac{6\Gamma^2(-\ep)}{\Gamma(3-2\ep)}
(-z)^{\ep}\right] + O(z).
\end{equation}
The application of Lagrange's method of finding a particular 
solution gives
\begin{equation}
J_{3p}^{(d)}=\frac{3z m^{2d-6}}{(1+\sqrt{z})^2}
\Gamma^2\left(1-\frac{d}{2}\right)
F_2\left(1,\frac12,\frac{d-1}{2},\frac{d}{2},d-1;\sqrt{z}R,R\right),
\label{J3pQLarge}
\end{equation}
where
\begin{equation}
R = \frac{4\sqrt{z}}{(1+\sqrt{z})^2},
\end{equation}
$F_2$ is the Appell function \cite{ApKdF} defined as
\begin{equation}
F_2(\alpha,\beta,\beta',\gamma,\gamma';x,y)=
\sum_{k,l=0}^{\infty}\frac{(\alpha)_{k+l}(\beta)_k (\beta')_l}
{(\gamma)_k (\gamma')_l} \frac{x^k ~y^l}{k!~l!}, ~~~|x|+|y|<1,
\end{equation}
and $(a)_n = \Gamma(a+n)/\Gamma(a)$ denotes the Pochhammer symbol.
Collecting all contributions, setting $d=4-2\ep$, applying
Euler's transformation for the first $\left._2F_1\right.$ function 
in (\ref{homoLarge}) we obtain the following solution of the
difference equation (\ref{j3difference}): 
\begin{eqnarray}
J_3^{(d)}&=&
 \frac{6 \Gamma^2(-\varepsilon) \Gamma^2(1+\varepsilon)
           (-z)^{\varepsilon} (1-z)^{ 2-2\varepsilon }}
 {m^{4\varepsilon-2}~ \Gamma(3-2\varepsilon) (1+3z)}
 \Fh21\FR{\frac13,\frac23}{2-\ep}
\nonumber \\
&+& \frac{\Gamma(-1+2\varepsilon)\Gamma^3(1-\varepsilon)
  (-z)^{2\varepsilon}
  (1-9z)^{2-2\varepsilon }}
  {m^{4\varepsilon-2}~\Gamma(3-3 \varepsilon)z(1+3z)}
 \Fh21\FR{\frac13,\frac23}{\ep}
\nonumber \\ 
&+&
\frac{3z m^{2-4\varepsilon}}{(1+\sqrt{z})^2}
\Gamma^2\left(-1+\varepsilon \right)
F_2\left(1,\frac12,\frac32-\varepsilon,
                   2-\varepsilon ,3-2\varepsilon;\sqrt{z}R,R\right).
\label{J3QLarge}
\end{eqnarray}
This is our main result.

The imaginary part of $J_3^{(d)}$ on the cut comes from
the first two terms in  (\ref{J3QLarge}). 
The analytic continuation of the two $\left._2F_1 \right. $ functions 
gives the relatively simple expression:
\begin{equation}
{\rm Im} J_3^{(d)} = \frac{-4 z~\pi^2 \sqrt{3\pi}
m^{2-4\ep}}
{\Gamma\left(\frac{3}{2}-\ep\right) \Gamma\left(2-\ep
\right)(1+3z)}
\left[\frac{(1-9z)^2}{108z^2}\right]^{1-\ep}
 \Fh21\FX{\frac13,\frac23}{2-\ep}.
\end{equation} 

At $d=4$ for the imaginary part this verifies the
result of \cite{BBBB}. 
Expanding  $J_3^{(d)}$ at $q^2=9m^2$  we reproduced 
the singular and finite in $\ep$  parts of  several  
terms of the on-threshold asymptotic expansion 
presented in Ref. \cite{Davydychev:1999ic}.

We found the following integral representation for Appell's  
$F_2$ function in (\ref{J3QLarge})
\begin{eqnarray}
&&
F_2\left(1,\frac12,\frac32-\varepsilon,
                   2-\varepsilon ,3-2\varepsilon;\sqrt{z}R,R\right)
\nonumber \\
&&~~~~~~
=
\frac{2 \Gamma(3-2\ep)}{\Gamma^2\left(\frac32-\ep\right)}
(1+\sqrt{z})^2~
\int_0^1
\frac{dt~ [t(1-t)]^{\frac12-\ep}}{(4zt+1-z+L)}
~\Fh21\Fus{1,\ep}{2-\ep},
\end{eqnarray}
where
\begin{equation}
L=\sqrt{(4zt-1-z)^2-4z}  ~.
\end{equation}
This integral representation can be used for the $\ep$ expansion of
the $F_2$ function. Integral representation for the $\left._2F_1\right.$ 
functions convenient  for  $\varepsilon$  expansion 
is given in the Appendix.

The analytic continuation of $J_3^{(d)}$ valid near the singular
points $q^2=0,m^2,9m^2$ can be directly  obtained from (\ref{J3QLarge})
by performing the  analytic continuation of the hypergeometric 
functions. Explicit formulae of the analytic continuations of $J_3^{(d)}$ 
in terms of Olsson functions \cite{Olsson} as well as the
$\ep$ expansion of the result we are planning to present in
a separate publication \cite{Tarasov2006}.

Using (\ref{J3QLarge}) we can find the on-threshold value of the 
integral. In formula (\ref{J3QLarge}) at $q^2=9m^2$ the imaginary 
part of the first term cancels the imaginary part of the second term. 
The  real part  coming from the $\left._2F_1\right.$
terms cancels the term with $\left._2F_1\right.$ which comes from
the Appell function $F_2$  at the threshold
\begin{eqnarray}
&&F_2\left(1,\frac32-\ep,\frac12,3-2\ep,2-\ep;\frac34,\frac14
     \right)=
\nonumber \\
&&~~~~+\frac{16(1-\ep)}{3(1-2\ep)}
 \Fh32\Fot{1,-1+2\ep,\frac32-\ep}{\frac12+\ep,2-\ep}
\nonumber \\
&&~~~~+\frac{4\Gamma(3-2\ep)\Gamma(2-\ep)
\Gamma\left(-\frac12+\ep\right)\Gamma(2-2\ep)}
{3^{\frac32-\ep}~\Gamma^2\left(\frac32-\ep\right)
\Gamma\left(\frac52-2\ep\right)}
 \Fh21\Fot{-\frac12+\ep,2-2\ep}{\frac52-2\ep}.
\end{eqnarray}
The cancellation of the $\left._2F_1\right.$ functions 
happens due to the fact that
\begin{equation}
\Fh21\Fot{2-2\ep,-\frac12+\ep}{\frac52-2\ep}
=\frac{2^{\frac52-2\ep}}{3^{\frac32-\ep}}
\frac{\Gamma\left(\frac54-\ep\right)\Gamma\left(\frac74-\ep\right)}
{\Gamma\left(\frac43-\ep\right)\Gamma\left(\frac53-\ep\right)}.
\end{equation}
Adding contributions from different hypergeometric functions
gives a rather simple expression for the  diagram at $q^2=9m^2$
\begin{eqnarray}
\left.J_3^{(d)}\right|_{q^2=9m^2}&=&
\frac{\Gamma^2(\ep)}{(1-\ep)(1-2\ep)}
 \Fh32\Fot{1,-1+2\ep,\frac32-\ep}{\frac12+\ep,2-\ep}
\nonumber
\\
&=&\frac{\Gamma^2(1+\ep)}{(1-\ep)(1-2\ep)}
\left\{-\frac{3}{2\ep^2}+\frac{9}{4\ep}
+\frac{75}{8} - \frac{8\pi}{\sqrt{3}}+ O(\ep) \right\}.
\label{on-threshold}
\end{eqnarray}
The first several terms in the $\ep$ expansion are in agreement with
the result of Ref.\cite{Davydychev:1999ic}.

\section{Conclusions}

Here we would like to add several remarks, which underline
the most important points which follow from the 
results of this paper.

Using a new method, for the first time, we were able to obtain 
an analytic expression for the two-loop sunrise, self-energy diagram
with equal mass propagators. Until now all attempts to find such 
a result  with other methods failed  for this integral.
This clearly demonstrates that the method  of dimensional 
recurrences is a  powerful tool for calculating Feynman integrals.
In our opinion there is a deep reason why, for example,
with the differential equation method \cite{Kotikov:1990kg} used in
\cite{Laporta:2004rb} an explicit formula could not be found. 
It turns out that even the associated  homogeneous  differential 
equation is rather complicated. Relevant for this case is
the  Heun equation \cite{Heun} 
with four regular  singular points,  located at $q^2=0,m^2,9m^2,\infty$. 
In general the reduction of the Heun equation to the hypergeometric
equation is a complicated mathematical problem \cite{Maier} 
which is not completely solved until now.

At the same time
the associated homogeneous  difference equation for $J_3^{(d)}$ 
is rather simple, and admits a reduction to a hypergeometric 
type of equation with linear  coefficients. 
 
 In fact this  is a rather general situation. Kinematical
singularities of Feynman integrals are located on 
complicated manifolds. In the case when the differential equations
are  of the first order there are no problems to solve them.
However, to solve a second or higher order differential equations
 in general will be  a problem because of complicated 
structure of the kinematical singularities.

The location of the singularities of Feynman integrals with
respect  to the  space time dimension $d$ is well known.
This has been used for  a rather evident rescaling of the integral by
ratios of $\Gamma$ functions which 
allowed us  to significantly reduce the order of the polynomial 
coefficients in the difference equation as we have seen in Sec. 3.
Finally this simplification allowed us to obtain the 
explicit result.

We expect that a further development of the method we used
in the present paper will help to find analytic results
for other more complicated Feynman integrals.

\section{Appendix}

The series representation for the Appell function $F_2$:
\begin{equation}
F_2(\alpha,\beta,\beta',\gamma,\gamma';x,y)=
\sum_{k=0}^{\infty}\frac{(\alpha)_{k}(\beta)_k}
{(\gamma)_k k!} ~x^k \left._2F_1(\alpha+k,\beta'; \gamma';y)\right.
\end{equation}
is convenient  for the analytic continuations  and also for 
the evaluation of $F_2$ at some particular values 
of their arguments.
The most frequently used integral representations for $F_2$ is
\begin{eqnarray}
&&F_2(\alpha, \beta, \beta', \gamma, \gamma'; w,z)=
\frac{\Gamma(\gamma) \Gamma(\gamma')}
{\Gamma(\beta) \Gamma(\beta') \Gamma(\gamma-\beta)
\Gamma(\gamma'-\beta')}
\nonumber \\
&&~~~~~~~\times \int_0^1 \int_0^1~du~dv~
u^{\beta-1}v^{\beta'-1} (1-u)^{\gamma-\beta-1}
(1-v)^{\gamma'-\beta'-1} (1-uw-vz)^{-\alpha}.
\end{eqnarray}

For the special parameters which appear in the explicit 
result  (\ref{J3QLarge}) we have:
\begin{eqnarray}
&&F_2\left(1,\frac32-\ep,\frac12,3-2\ep,2-\ep;x,y \right)
\nonumber \\
&&~~~~~~=\frac{\Gamma(3-2\ep)\Gamma(2-\ep)}
{\Gamma^3\left(\frac32-\ep \right)\Gamma\left(\frac12\right)}
\int_0^1 \int_0^1 \frac{dudv}{\sqrt{v}}
\frac{[u(1-u)(1-v)]^{\frac12-\ep}}{(1-uw-vz)}.
\end{eqnarray}

Furthermore, 
we found the following specific relation between the
$\left._2F_1\right.$   function and yet another Appell's  
function $F_1$ 
\begin{equation}
 \Fh21\FR{\frac13,\frac23}{2-\ep}
=\frac{(1+3z)}{(1-z)}
F_1\left( \frac12, -\frac12+\ep,-\frac12+\ep,2-\ep;
\frac{4z}{(1+\sqrt{z})^2},\frac{4z}{(1-\sqrt{z})^2}\right).
\end{equation}
To our knowledge such a relation has not been found 
sofar in the mathematical literature.
The  integral representation for the $F_1$ function reads
\begin{equation}
F_1\left( \frac12, -\frac12+\ep,-\frac12+\ep,2-\ep;w,z \right) =
\frac{\Gamma(2-\ep)}
{\Gamma\left(\frac12 \right) \Gamma\left(\frac32- \ep \right)}
\int_0^{1}\frac{du}{\sqrt{u}}[(1-u)(1-wu)(1-zu)]^{\frac12-\ep}
\end{equation}
and is convenient for the $\ep$ expansion.
This $F_1$ function can be considered as a generating function
of  a new generalization of elliptic integrals which may appear
in evaluating Feynman integrals.

\vspace{1cm}
{\bf Acknowledgments.}~~
I am very thankful to Fred Jegerlehner for useful remarks
and carefully reading the manuscript. 
This work was supported by DFG Sonderforschungsbereich 
Transregio 9-03. 

%%%%%%%%%%%%%%%%%%%%%%%%%%%%%%%%%%%%%%%%%%%%%%%%%%%%%%%%%%%%%%%%%%%%%%%%%%%%%%%

\end{document}